\documentclass[a4paper]{easychair}
\usepackage[british]{babel}
\usepackage[abbreviations,british]{foreign}
\usepackage[useregional]{datetime2}
\usepackage{glossaries}
\usepackage{microtype}
\usepackage{url}
\usepackage{xspace}
\usepackage{bera}% optional: just to have a nice mono-spaced font
\usepackage{listings}
\usepackage{xcolor}

\definecolor{eclipseStrings}{RGB}{42,0.0,255}
\definecolor{eclipseKeywords}{RGB}{127,0,85}
\colorlet{numb}{magenta!60!black}

\lstdefinelanguage{json}{
    basicstyle=\normalfont\ttfamily,
    commentstyle=\color{eclipseStrings}, % style of comment
    stringstyle=\color{eclipseKeywords}, % style of strings
    numbers=left,
    numberstyle=\scriptsize,
    stepnumber=1,
    numbersep=8pt,
    showstringspaces=false,
    breaklines=true,
    frame=lines,
    string=[s]{"}{"},
    comment=[l]{:\ "},
    morecomment=[l]{:"},
    literate=
        *{0}{{{\color{numb}0}}}{1}
         {1}{{{\color{numb}1}}}{1}
         {2}{{{\color{numb}2}}}{1}
         {3}{{{\color{numb}3}}}{1}
         {4}{{{\color{numb}4}}}{1}
         {5}{{{\color{numb}5}}}{1}
         {6}{{{\color{numb}6}}}{1}
         {7}{{{\color{numb}7}}}{1}
         {8}{{{\color{numb}8}}}{1}
         {9}{{{\color{numb}9}}}{1}
}
\usepackage{bera}% optional: just to have a nice mono-spaced font
\usepackage{listings}
\usepackage{xcolor}

\definecolor{eclipseStrings}{RGB}{42,0.0,255}
\definecolor{eclipseKeywords}{RGB}{127,0,85}
\colorlet{command}{magenta!60!black}

\lstdefinelanguage{cmd}{
    basicstyle=\normalfont\ttfamily,
    commentstyle=\color{eclipseStrings}, % style of comment
    stringstyle=\color{eclipseKeywords}, % style of strings
    keywordstyle=\color{eclipseKeywords}, % style of keywords
    numbers=left,
    numberstyle=\scriptsize,
    stepnumber=1,
    numbersep=8pt,
    showstringspaces=false,
    breaklines=true,
    frame=lines,
    string=[s]{"}{"},
    comment=[l]{rem},
    morekeywords={
        set,
        call
    }
}

\newcommand{\pp}{+\nolinebreak\hspace{-.2em}+}
\newcommand{\cpp}{C\nolinebreak\hspace{-.1em}\pp\xspace}
\newcommand{\opp}{\texorpdfstring{OMNeT\nolinebreak\pp}{OMNeT++}\xspace}

\newacronym{ide}{IDE}{Integrated Development Environment}
\newacronym{json}{JSON}{JavaScript Object Notation}
\newacronym{voip}{VoIP}{Voice over IP}
\newacronym{vsc}{VSC}{Visual Studio Code}

\title{Towards a modern CMake workflow}
\author{
    Heinz-Peter Liechtenecker\inst{1}
    \and
    Raphael Riebl\inst{2}
}

\institute{
    TU Graz, Institut für Elektronik, Inffeldgasse 12, 8010 Graz, Austria
    \and
    TH Ingolstadt, Esplanade 10, 85049 Ingolstadt, Germany
}

\authorrunning{Liechtenecker and Riebl}

\titlerunning{Towards a modern CMake workflow}

\begin{document}
\maketitle

\begin{abstract}
    Modern CMake offers the features to manage versatile and complex projects with ease.
    With respect to \opp projects, a workflow relying on CMake enables projects to combine discrete event simulation and production code in a common development environment.
    Such a combination means less maintenance effort and thus potentially more sustainable and long-living software.
    This paper highlights the significant improvements since the first attempt of using CMake in \opp projects.
    In particular, a state-of-the-art integration of \opp in Visual Studio Code including support for debugging and multi-platform compilation is presented.
    Last but not least, an exemplary use case demonstrates the powerful mix of production and simulation code in a common software architecture supported by the \opp CMake package.
\end{abstract}

\section{Introduction}
\label{sec:introduction}
Modular software architectures are of great help in designing platform-independent and long-living software, which allows developers to reuse components.
Furthermore, complex functionality can be broken down into smaller chunks to ease handling and testing.
In the context of event-driven simulation, it is desirable to reuse features such as protocol decoders or application logic:
Besides reduced development effort, the resulting simulation model closely resembles the application as found in production environments.
For example, the INET Framework relies on some FFmpeg libraries for its \gls{voip} feature instead of reinventing the wheel.

A well-designed software architecture and build environment reduces the management overhead and thus allows even small teams to maintain and improve complex projects.
CMake is a battle-proven tool supporting developers to create such a powerful build environment.
Modern CMake~\cite{CMake} can, among others, be employed for the management of various build targets, automation of pre- and post-build tasks, and cross-platform toolchain orchestration.
CMake can generate a native build environment that compiles source code, creates libraries, generates wrappers, and builds executables in arbitrary combinations.
CMake supports in-place and out-of-place builds, and can therefore build multiple variants from a single source tree~\cite{CMakeAbout}.
Furthermore, CMake is capable of generating various build contexts, ranging from classic Makefiles to the more recent Ninja~\cite{NinjaBuild} build system, allowing CMake to instantiate a build environment across platforms and users.
Due to these facts, CMake has been widely adopted in various projects and libraries to allow for seamless cross-platform integration.

CMake neither interacts directly with the actual toolchains nor replaces them; it acts as an orchestrator for various tasks during the configuration and build process of the actual artefact.
CMake and the package presented within this paper complements \opp. As soon as \opp itself has been set up, one can switch over from the \opp build system to CMake, which introduces, among others, the following advantages:
\begin{itemize}
    \item Simplified dependency management
    \item Integration of CMake-based projects and libraries
    \item Simplified management of multiple build targets (e.g. production, simulation, and test targets) from a single source tree
    \item Projects employing CMake can easily be integrated in a variety of \glspl{ide}
\end{itemize}

Usage of CMake as an option to handle software dependencies in \opp simulations has first been presented at the \opp Community Summit in 2015~\cite{Riebl2015a}.
Maintenance of the presented tooling has not ceased since then; it took place within the Artery project, however.
This intermingling with Artery made it unnecessarily hard for other projects to adopt CMake as well.
Meanwhile, Thor K. Høgås extracted the related sources into a separate repository in 2020.
Since 2021, this repository has moved under the umbrella of the \opp organisation on GitHub\footnote{\url{https://github.com/omnetpp/cmake}}.
Thor and the authors of this paper are now jointly maintaining this standalone \opp CMake package for (hopefully) wide usage by the \opp community.

Beyond these organisational changes, this \opp CMake package has also been drastically improved in terms of features.
This paper highlights some of these improvements:
Section~\ref{sec:opp-deps} revisits how to deal with \opp projects in CMake and how the related mechanisms have changed.
Furthermore, we outline in Section~\ref{sec:cmake-ides} how \opp projects can be developed with alternative \glspl{ide} such as \gls{vsc}.
In particular, we bring convenient features such as \opp debug targets to \gls{vsc}.
Section~\ref{sec:cmake-targets-ci} shows an exemplary project structure, which leverages CMake to have build targets for production and simulation software in a common setup.
Such a setup makes it easy to run continuous integration tests in a simulation environment to strengthen confidence in the software executed in a production environment.

\section{Depending on \opp Projects}
\label{sec:opp-deps}
One of CMake's key selling points is its ability to handle third-party software:
It is a piece of cake for developers to add such dependencies to their projects -- if someone has extended CMake's \emph{find} capabilities accordingly.
With the help of our \opp CMake package, \texttt{find\_package(OmnetPP)} is readily available which provides \emph{import targets} for all \opp libraries and \emph{variables} to \opp tools such as the message compiler.

If a project depends not only on \opp itself but also intends to reuse models by \opp frameworks such as INET or SimuLTE, these frameworks need to be covered by CMake as well.
In the \opp CMake package's first incarnation, a project had to use the \texttt{import\_opp\_target} macro to add another \opp project as its dependency.
As detailed in~\cite{Riebl2015a}, the \emph{Makefile} produced by \texttt{opp\_makemake} is parsed by our \texttt{opp\_cmake} script generating an adequate CMake \emph{import target} out of it.
However, the original build process of the \opp project was still in charge of creating the project's binaries via its \emph{Makefile}.

Nowadays, \texttt{add\_opp\_target} provides an alternative mechanism:
Instead of merely importing binaries built by an external toolchain, \texttt{add\_opp\_target} lets CMake itself build the \opp project's binaries.
It can not only handle the \cpp source files (\emph{*.cc}) but also \opp messages (\emph{*.msg}), \ie{} it invokes the \opp message compiler as an intermediate build step.
Similar to its \texttt{import\_opp\_target} sibling, the resulting target has its \emph{NED\_FOLDERS} property configured.
Hence, the project's NED paths are automatically considered when a depending simulation model is run.

The main advantage of \texttt{add\_opp\_target} over \texttt{import\_opp\_target} is the consistent build chain.
A CMake project employing \texttt{import\_opp\_target} has to trigger the execution of the \opp Makefile, even if the CMake project itself employs another build generator such as Ninja.
With \texttt{add\_opp\_target}, all build steps are fully controlled by CMake itself.
Hence, one can easily exploit the speed-up enabled by parallelised Ninja builds.
Furthermore, the compiler options used for \opp build artefacts are entirely configurable via the common CMake mechanisms.

\section{Alternative IDEs}
\label{sec:cmake-ides}

The official \opp \gls{ide} is the natural choice for many \opp users as it neatly integrated \opp specifics.
Especially novice users thus do not have to worry about peculiarities such as setting up NED paths.
As long as the simulation models remain rather self-contained, \ie{} no dependencies to third-party components exist, the workflow in the \opp \gls{ide} is rather straightforward.

As soon as users head for more sophisticated setups, our \opp CMake package can help out to tackle the management of dependencies as pointed out in Section~\ref{sec:opp-deps}.
Unfortunately, the built-in \opp \gls{ide} plugins cannot handle such a CMake environment as conveniently as the classic \emph{opp\_makemake} projects.
However, a CMake-based \opp project can be handled by any \gls{ide} with decent CMake support.
In the following, we explore the state-of-the-art in developing \opp projects with \gls{vsc}~\cite{VisualStudioCode}.

A modern and robust \gls{ide} has to fulfil several requirements, including code highlighting and syntax checks, auto-completion and suggestions, integration with version control systems, execution of builds, debugging and execution and analysis of tests.
In recent years, \gls{vsc} has become a popular choice for a plethora of use cases.
Customisability and cross-platform support are key features of \gls{vsc}, which allow us to neatly adopt it for \opp development as well.
Especially with its extensions to build and debug C/\cpp code managed by CMake, \gls{vsc} is an ideal candidate to work with the \opp/CMake ecosystem.
The authors suggest the following extensions making \gls{vsc} a fully-featured \gls{ide}:
\begin{itemize}
    \item \emph{ms-vscode.cpptools} for C/\cpp language support~\cite{VSC-CppTools}
    \item \emph{twxs.cmake} for CMake language support~\cite{VSC-CMakeLanguageSupport}
    \item \emph{ms-vscode.cmake-tools} for CMake project integration~\cite{VSC-CMakeTools}
    \item \emph{schrej.omnetpp-ned} for \opp NED language support~\cite{VSC-NedLanguageSupport}
    \item \emph{vadimcn.vscode-lldb} (CodeLLDB) for debugging \cpp applications~\cite{VSC-CodeLLDB}
\end{itemize}

\subsection{Debugging \opp Models}
Besides code editing, debugging is a major requirement within any \gls{ide}.
For launching an actual debug session, \gls{vsc} depends on additional instructions given in the \gls{vsc}-specific
\textit{launch.json} file~\cite{VSC-Debugging}.
In particular, debugging an \opp model requires instructions about the desired debugger, the location of the \opp runner executable (usually \emph{opp\_run\_debug}), the \opp model and associated command-line parameters.
These command-line parameters specify the look-up paths for \emph{NED} files, the to be used simulation configuration file (an \emph{omnetpp.ini} in most cases), and the \opp model's libraries.
Managing these parameters manually is a tedious and error-prone task, especially for large and complex projects.
For example, Artery configurations often employ multiple other \opp models, \eg{} SimuLTE and INET, plus its own modules.

As a remedy for this maintenance burden, recent versions of the \opp CMake package can configure the \gls{vsc} debug targets automatically.
The build system then updates the debug configuration in the \textit{launch.json} file according to the CMake project configuration.
If this feature is enabled, \gls{vsc} debug configurations are generated for every \texttt{add\_opp\_run} invocation.
As before, \texttt{add\_opp\_run(<name>)} still adds \emph{run\_<name>} (always), \emph{debug\_<name>} (for debug builds), and \emph{memcheck\_<name>} (if Valgrind is installed) targets.
Now, it also invokes a Python script shipped with the \opp CMake package that updates the \textit{launch.json} in the project's \textit{.vscode} directory.
Manual configurations by the user remain untouched, though.
Hence, it is perfectly fine to have generated and custom debug configuration side-by-side as long as their names are distinct.

At the moment, two flavours of debug configurations are supported: GDB~\cite{GDB-GnuDebugger} and LLDB~\cite{LLDB-LLVMDebugger}, the debuggers by the GNU and LLVM projects, respectively.
Listing~\ref{lst:launchjson} shows an exemplary configuration for the \gls{vsc} CodeLLDB extension employing the LLDB debugger.
We have found this extension to be advantageous for debugging across platforms, \ie{} on Windows and Linux systems.
The upcoming \opp 6.0 release will also ship with an \opp formatter script for the LLDB debugger.
For these combinations, \ie{} CodeLLDB and an \opp 6.0 pre-release, the generated debug configurations load this formatter script as part of the debugger's initialisation commands.
Thus, the debugger pretty-prints the \opp data structures out-of-the-box then.

\begin{lstlisting}[language={json}, caption={\Gls{vsc} debug configuration using CodeLLDB (\emph{launch.json})}, label={lst:launchjson}]
{
    "version": "0.2.0",
    "configurations": [
        {
            "name": "Launch MyProject - CodeLLDB (OMNeT++)",
            "type": "lldb",
            "request": "launch",
            "program": "path/to/bin/opp_run_dbg",
            "args": [
                "-n", "path/to/ned/folders",
                "-l", "path/to/library",
                "path/to/omnetpp.ini"
            ],
            "stopOnEntry": false,
            "cwd": "path/to/working/directory",
            "initCommands": [
                "command script import path/to/lldb/formatters/omnetpp.py"
            ]
        }
    ]
}
\end{lstlisting}

Unfortunately, the Python's native \gls{json} package does not support C-style comments within \gls{json} files and is not as fault-tolerant as the \gls{vsc}'s \gls{json} decoder.
Hence, when manually editing the \textit{launch.json} file, the user must ensure to comply strictly with the \gls{json} standards for now.
We may be able to lift this limitation in the future.

\subsection{Improving Developer Experience on Windows with \texorpdfstring{\gls{vsc}}{VSC}}
\label{sec:cmake-windows}

\Gls{vsc} and the corresponding CMake extension allows to set up a modern and flexible cross-platform development environment.
Especially developing \opp packages and simulations on Windows using alternatives to the shipped Eclipse-based \gls{ide} can often be error-prone as \opp comes with its own \cpp toolchain environment.
Integrating the virtual environment in alternative \glspl{ide} like \gls{vsc} in conjunction with the \opp CMake can be achieved quickly - for \gls{vsc}, the CMake extension needs to be aware of the shipped \opp compilers.
In addition, the virtual environment binary paths have to be part of the systems PATH environment variable.
Paths can be either added globally or by executing an environment setup script.
The \gls{vsc} CMake extension offers the necessary feature: Utilising user local kits, one can add additional toolchains within a \textit{cmake-kits.json} file~\cite{VSC-CMakeKits}.
Listing~\ref{lst:cmakekitsjson} gives an example of a \textit{cmake-kits.json} file setting up a kit for \opp 6.0pre10 using the Clang compiler and executing a command script to set up the environment.
The environment setup script, shown in Listing~\ref{lst:envsetup}, sets the PATH variable and optionally activates a virtual python environment.

\begin{lstlisting}[language={json}, caption={Example of a \gls{vsc} CMake Kits configuration (\textit{cmake-kits.json})}, label={lst:cmakekitsjson}]
[
  {
    "name": "CLang OMNeT++ 6.0pre10 with Python VENV",
    "environmentSetupScript": "${workspaceFolder}/.vscode/omnetpp-6.0pre10env.cmd",
    "compilers": {
      "C": "path/to/omnetpp-6.0pre10/tools/win64/mingw64/bin/clang.exe",
      "CXX": "path/to/omnetpp-6.0pre10/tools/win64/mingw64/bin/clang++.exe"
    }
  }
]
\end{lstlisting}

\begin{lstlisting}[language={cmd}, caption={Environment setup script (\textit{omnetpp-6.0pre10env.cmd})}, label={lst:envsetup}]
set PATH=%PATH%;path\to\omnetpp-6.0pre10\tools\win64\mingw64\bin
set PATH=%PATH%;path\to\omnetpp-6.0pre10\bin
set PATH=%PATH%;path\to\omnetpp-6.0pre10\tools\win64\opt\mingw64\bin
set PATH=%PATH%;path\to\omnetpp-6.0pre10\tools\win64\usr\bin
set PATH=%PATH%;path\to\omnetpp-6.0pre10\lib

rem Optional: Activate a python virtual environment
set current_dir="%~dp0"
call %current_dir%..\.venv\Scripts\activate.bat
\end{lstlisting}

\begin{figure}[h]
    \begin{centering}
        \includegraphics[width=\textwidth]{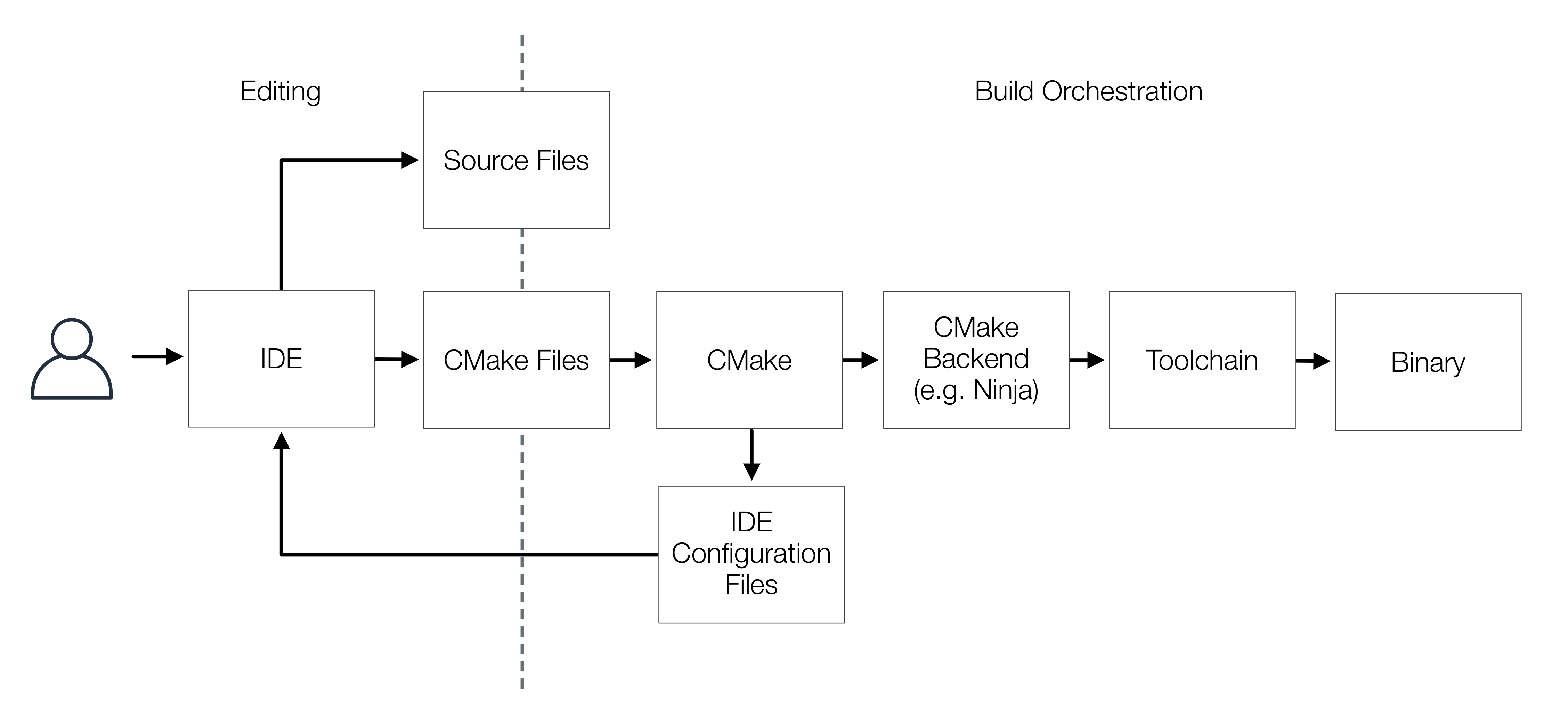}
        \caption{CMake Toolchain and \gls{ide} integration}
        \label{fig:cmake_toolchain}
    \end{centering}
\end{figure}

Taking full advantage of the presented features, developing and testing simulations and packages against different \opp versions becomes possible by simply switching the CMake kit within \gls{vsc}.
Figure~\ref{fig:cmake_toolchain} gives an overview of the \gls{ide} experience:
Besides building binaries, CMake also configures the \gls{ide} itself based on the given build-targets.

\section{Use Case: Build Targets and Continuous Integration}
\label{sec:cmake-targets-ci}

Various factors lead to the necessity to reuse portions of source code within simulation and production environments.
Especially for small teams, reuse is a vital aspect so projects can be maintained for years.
With suitable software and build architecture in place, improvements found by simulation can be directly integrated into production code.
Reducing redundancy between simulation and production code is a critical factor for development efficiency.
In conjunction with a suitable build management tool such as CMake, build targets tailored to production can easily integrate existing source code.
Typical build targets may comprise:
\begin{itemize}
    \item Dynamically loaded libraries to run a simulation
    \item Test executables putting various code sections through their paces
    \item A standalone executable reusing parts of the code used for simulation within a production environment
\end{itemize}

\begin{figure}
    \begin{centering}
        \includegraphics[width=\textwidth]{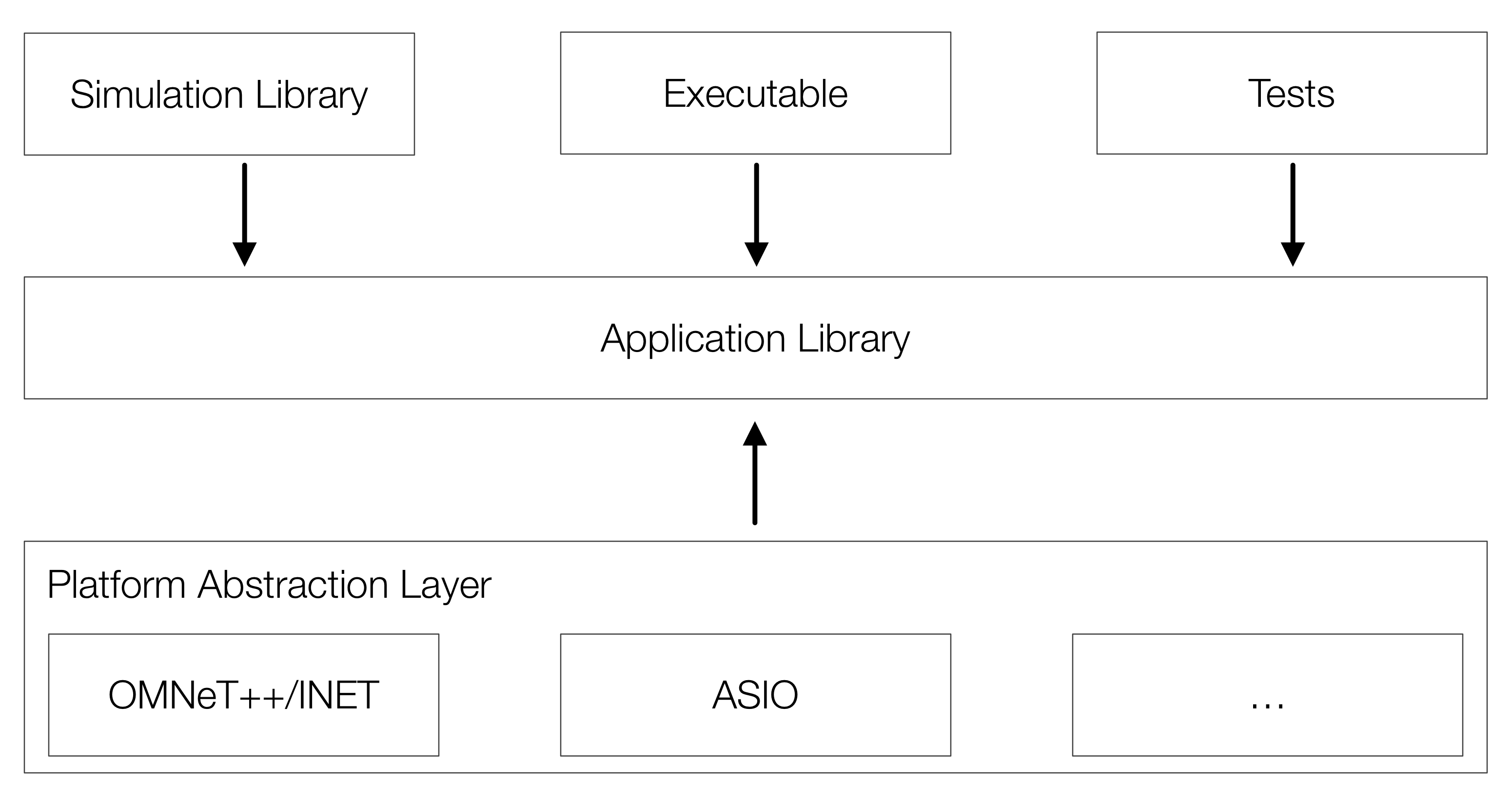}
        \caption{CMake targets and their dependencies in a mixed simulation \& production project}
        \label{fig:cmake_overview}
    \end{centering}
\end{figure}

Figure~\ref{fig:cmake_overview} shows an exemplary project structure featuring a \emph{simulation library} with \opp modules, a \emph{standalone executable} for productive operation, and \emph{tests}.
In the given example, a network application shall receive TCP traffic, process it and respond to it if necessary.
The \emph{application library} includes this core functionality, which is employed by all three types of build targets as mentioned above.
The \emph{platform abstraction layer} provides the networking functionality required by the application library via a platform-agnostic interface:
From the application's perspective, it does not matter if the network is simulated or an actual network is accessed.
The respective implementations of this interface can rely on \opp/INET for a simulated runtime context or the ASIO library, which employs the operating system's network stack.
ASIO~\cite{ASIO-thinkasync} and INET~\cite{INET-framework}, in particular its socket implementation, behave both like event-driven I/O systems.
With this common behaviour of INET sockets and the ASIO library, a generalised interface can easily be offered to the application library.
% TODO Should we add references for INET and ASIO? At least ASIO may not be well-known by the OMNeT++ community.

Two variants are apparent to realise the platform abstraction layer within CMake:
Either distinct (static or dynamic) library targets exist for each implementation, or all flavours are built into a single binary, \eg{} using the pointer-to-implementation idiom at runtime~\cite{Meyers2015a}.
With the former approach, each library depends only on some platform-specific libraries.
For example, the "real" implementation depends on the ASIO library, while the "simulation" implementation depends only on \opp and INET.
In any case, the platform variants are realised as CMake targets, and thus CMake takes care of all the (transitive) dependencies automatically.

Without changes at the application logic itself, the executable for productive operation can employ the ASIO-based platform while other components may use the \opp-based platform.
Results and improvements found by simulation and verification are directly contributed to a common code base represented by the application library.
This pattern predestines for a continuous integration pipeline.
Ideally, the build system is part of this pipeline to build code upon specific events, \eg{} a Git commit or pull request.
By dividing the application into several targets, it is possible only to build those of interest:
the application library with the corresponding platform abstraction implementation, the standalone and the
testing executables.
To sum up, the proposed CMake workflow avoids the management overhead occurring when separate build systems need to be maintained for each platform.

\section{Conclusion}
\label{sec:conclusion}
Due to the presented advances in the \opp CMake package, developers familiar with CMake workflows can fully integrate existing \opp packages like INET into their projects with ease:
Developers starting a new project relying on existing \opp packages do not have to port the Makefiles to CMake build instructions first.
If porting a package to CMake is desirable, for instance to allow for modularizing a complex \opp package or simplifying dependency integration, the developer needs to reassemble the existing Makefiles with equivalent CMake instructions; there is no need to change the actual source code, though.
CMake build instructions and Makefiles can also coexist until the CMake setup is considered stable enough to substitute the Makefiles ultimately.
The extended functionality of the CMake package to automate the setup of \gls{vsc} as a fully-featured integrated development environment improves the developer experience on Linux- and Windows-based systems.

Further work on the package includes optimisation of the environment setup when debugging simulations within \gls{vsc} on Windows. The approach presented in Section~\ref{sec:cmake-windows} currently only sets the environment variables for the build steps; hence debugging requires adding \opp paths to the global Windows system \textit{PATH} variable.
Furthermore, adding examples to show the  \opp CMake package usage, a project skeleton, and improved documentation is necessary to lower the entry bar for developers heading towards a modern CMake workflow with \opp.

\newpage
\bibliography{references.bib}
\bibliographystyle{ieeetr}

\end{document}